# Two-phase flow: structure, upscaling, and consequences for macroscopic transport properties

R. Toussaint[1,6], K.J. Måløy[2,6], Y. Méheust[3,6], G. Løvoll[2, 4], M. Jankov[2], G. Schäfer[4], J. Schmittbuhl[1]

(1) IPGS, CNRS, University of Strasbourg, Strasbourg (France)
(2) Physics Department, University of Oslo, Oslo (Norway)
(3) Géosciences Rennes, University of Rennes 1, Rennes (France)
(4) Det Norske Veritas AS, Research and Innovation, Høvik (Norway)
(5) LHYGES, CNRS, University of Strasbourg, Strasbourg (France)
(6) Centre for Advanced Study, Centre for Advanced Study at The Norwegian Academy of Science and Letters, Oslo (Norway)

January 13, 2012

**Abstract**

In disordered porous media, two-phase flow of immiscible fluids (biphasic flow) is organized in patterns that sometimes exhibit fractal geometries over a range of length scales, depending on the capillary, gravitational and viscous forces at play. These forces, as well as the boundary conditions, also determine whether the flow leads to the appearance of fingering pathways, i.e., unstable flow, or not. We present here a short review of these aspects, focusing on drainage and summarizing when these flows are expected to be stable or not, what fractal dimensions can be expected, and in which range of scales. We base our review on experimental studies performed in two-dimensional Hele-Shaw cells, or addressing three dimensional porous media by use of several imaging techniques. We first present configurations in which solely capillary forces and gravity play a role. Next, we review configurations in which capillarity and viscosity are the main forces at play. Eventually, we examine how the microscopic geometry of the fluid clusters affects the macroscopic transport properties. An example of such an upscaling is illustrated in detail: For air invasion in a mono-layer glass-bead cell, the fractal dimension of the flow structures and the associated scale-ranges, are shown to depend on the displacement velocity. This controls the relationship between saturation and the pressure difference between the two phases at the macroscopic scale. We provide in this case expressions for dynamic capillary pressure and residual fluid phase saturations.

# 1 Introduction

The physics of two phase flows in porous media is a complex and rich topic, with obvious applications to the hydraulics of the vadoze zone, be it water infiltration, its evaporation, or the transport of Dense Non-aqueous Phase Liquids (DNAPL) down to the aquifers (Dridi et al., 2009). Hydrogeologists and soil scientists aim at relating volumetric flow, pressure head, and water content at the Darcy scale, which is a meso-scale above which the medium and the flow are described by continuous mathematical fields. They also need to predict the front displacement of the injected fluids, its localizing or non localizing character, and the fluid mass distribution behind it. The basic laws of multiphase flows treated at mesoscopic scale as a continuum require a closure of simultaneous flow according to Darcy's law. A key point of this closure is a functional relation between the capillary pressure and (water) saturation in the form of retention curves; another key point in the dependence of the relative permeabilities on saturation.

The physics community has been mostly concerned with characterizing and understanding flow structures/patterns from the pore scale and up. These structures and processes have a major impact on the retention curves (see e.g. review by Blunt (2001)). Notably, viscous fingering may have strong influence on retention curves, resulting in dynamic saturation–pressure curves in porous media as we will show in section 4 below.

These flow structures can vary from compact to ramified and fractal (Lenormand et al., 1988; 1989; Måløy et al., 1985; Méheust et al., 2002; Sandnes et al., 2011; Holtzman and Juanes, 2010). One major issue is to simplify this complexity by keeping just enough information to describe the relevant physics at the relevant scale for the flow considered, without discarding essential information. For example, the simple invasion percolation model is sufficient to model the flow structure obtained under slow drainage conditions.

The fact that such simple models can describe simple features of complex systems arises from the property of universality in critical states: many physical dynamic systems, governed by a competition of simple forces with a disorder in thresholds, are in the vicinity of so-called critical points in statistical physics, as discussed e.g. by Domb (1996) or Feder (1988): these are characterized by scale invariance over some ranges, fractal dimensions, and by critical exponents precising how characteristic lengths that limit the fractal ranges depend on the system size or driving speed. An interesting aspect of such critical points is that the corresponding exponents and fractal dimensions do not depend on the small scale details of the system, but are controled by how the systems are invariant under some form of upscaling. Consequently many systems differing at small scale are characterized by the same critical exponents: they are said to belong to the same universality class. This allows to describe complex systems using sometimes simple computer models.

The porous body of a piece of soil or rock consists of pores and fracture networks of different length scales and shapes, whose permeability presents large spatial variations. These structures can be correlated at large scale (see e.g. Brown, 1995; Zimmerman and Bodvarsson, 1996; Méheust and Schmittbuhl 2003, Neuville et al. 2010a) or present a finite correlation length (Neuville et al. 2010b). The variations in permeability results in flow channeling (see e.g. Brown, 1987; 1995, Drazer and Koplik, 2002, Méheust and Schmittbuhl 2000, 2001, 2003, Neuville et al., 2010a, 2011a, 2011b, 2011c) and a potential permeability anisotropy (Méheust and Schmittbuhl 2000, 2001). In general the soil/rock is a dynamic medium where the porosity can be modified by the fluids involved due to chemical reactions and desorption/adsorption mechanisms (Szymczak and Ladd, 2011), in addition to the fluid pressure and the mechanical stress acting on the porous medium (Johnsen et al., 2006; Goren et al., 2010; 2011). The chemical composition and nano/micro structure of the rock further decides the wetting properties of the fluids which is crucial for the capillary front advancement in two-phase flow. For example, when a fluid with high viscosity is displaced in a porous medium by a fluid with a lower viscosity, the displacing fluid tends to channel through the paths of lower flow resistance, thereby forming pronounced fingers. The physical properties of the fluids play an important practical role on natural flows: e.g., in soil and groundwater, the identification of pollution sources is difficult due to the fact that organic pollutants can rapidly migrate down to the bottom of the aquifer and/or along paths different from the water (Benremita and Schäfer, 2003).

In addition, when the porous medium is deformable, branching structures can be observed with transitions to fracturing of the porous medium (Lemaire et al., 1993; Cheng et al., 2008; Sandnes et al., 2011; Holtzman and Juanes, 2010; Chevalier et al., 2009; Johnsen et al., 2006; Varas et al., 2011) or formation of fingers, channels or bubbles in it (Johnsen et al., 2006; 2007; 2008; Kong et al., 2011; Vinningland et al., 2007a; 2007b, 2010, 2011; Niebling et al., 2010a; 2010b).

In this review we will address the detailed structure and dynamics of two phase flow in fixed and disordered porous media based on pore scale experiments. We will limit the discussion mostly to drainage, i.e. to situations where a non wetting fluid displaces a wetting one - even though imbition, where a wetting fluid invades a non wetting one, is of equal practical importance. The discussion will be limited to media that are isotropic and homogeneous at large scales, and to cases where chemical reactions and adsorption or desorption between the fluids and the porous medium can be neglected. The structures of clusters of the moving fluid and the dynamics of drainage in porous media depend on several parameters like the density difference, the surface tension, the wetting properties, the viscosities and the flow rates of the fluids involved. The various forces at play dominate on different length scales and their interplay give rise to separate scaling regimes. Up-scaling, which consists in relating the pore scale description to properties defined at the Darcy scale or even at the macroscopic scale, is a central topic within hydrology and petroleum engineering.

Only by understanding the scaling of the structures and dynamics within each regime, and the crossover lengths involved, it is possible to perform up-scaling. The structures involved are typically fractal within some scaling range; their fractal dimension depends on length scales, and often result from fluctuations occurring at smaller scales. At the end of this short review we provide an example of upscaling of recent experimental data; the experiments in question were aimed at studying the crossover between capillary and viscous fingering in a quasi two–dimensional (monolayer) porous medium.

This review summarizes the results of a series of works, published mostly in Physics journals, that are of interest to model flow in the vadose zone, or in general in hydrology. We will also illustrate on simple examples what type of microstructure and what properties of fingering of the flow control upscaling and the dynamic dependence of macroscopic capillary pressure on microscopic flow.

## 2  Capillary and gravitational effects

When drainage is performed in the limit of infinitely slow displacement velocities, the pressure drop accross the porous medium is controlled by the capillary pressure drop accross the interface between the two fluids. The criterion for advancement of the interface into a given pore is that the capillary pressure drop is larger than the capillary pressure threshold needed to invade the pore neck that separates that pore from the already-invaded adjacent pore. The value of the capillary pressure threshold fluctuates from pore neck to pore neck, with a distribution function determined by the geometry of the porous medium. In the case of zero gravity or for a horizontal 2D porous medium, the next pore throat /neck to be invaded will be, among the pores that touch the interface, the one whose pore throat has the smallest capillary pressure threshold. This idea is the basis of the invasion percolation algorithm (de Gennes and Guyon, 1978; Chandler et al., 1982; Wilkinson and Willemsen, 1983) where random numbers representing the capillary pressure threshold values are distributed on a lattice and where the front is moved at each time step at the location along the interface corresponding to the smallest threshold value. The fact that the fluid front always moves at the most easily invaded pore neck and nowhere else is actually not always true in real flows, even if it is a good approximation. What drives the advancement of the front is the capillary pressure build-up in the fluid. The capillary pressure will not relax immediately after invasion of a new pore but is controlled by a back contraction of the fluid interface. This is the reason for the so called Haines jumps which may lead to invasion of several pores in one jump (Haines, 1930; Måløy et al., 1992; Furuberg et al., 1996).

When the displaced fluid is incompressible (or lowly compressible), trapping takes place. Trapping is very important in two dimensions (2D) (Wilkinson and Willemsen, 1983) but much less significant in three dimensions (3D). Experiments addressing capillary fingering in 2D model systems were first performed by Lenormand et al. (1988; 1989) who found a mass fractal dimension of the invaded structures equal to $D_c = 1.83$, which is consistent with the results of numerical simulations based on invasion percolation (in the version of the model allowing trapping of the invaded fluid) (Wilkinson and Willemsen, 1983). In 3D, several experiments have been performed (Chuoke et al., 1959; Paterson et al., 1984a; 1984b; Chen et al., 1992; Frette et al.1994; Hou et al., 2009; Mandava et al., 1990; Nsir et al., 2011; Yan et al., 2012). The fractal dimension found at small scale (between 2.0 and 2.6) is compatible with the dimension $D = 2.5$ found in three dimensional invasion percolation models (Wilkinson and Willemsen, 1983).

Even though the capillary fingering structure is fractal, in practice it is well described by a fractal dimension only within a window of length scales ranging from the pore size up to a crossover length on a larger scale. In the case where the density difference between the two fluids is different from zero, but where viscous forces are small compared to the others, this crossover length corresponds to a scale at which the capillary threshold fluctuations become of the same order of magnitude as the difference in hydrostatic pressure drop between the two fluids. This means that the crossover will always occur when the fluid structures become large enough. When a lighter fluid is displacing a heavier one from above at a slow flow rate resulting in low viscous forces, a stable displacement is observed. In this case, the displacing fluid does not finger its way through the displaced fluid; the crossover length sets the width of the rough interface between the two fluids (Birovljev et al., 1991; Méheust et al., 2002; Løvoll et al., 2005), parallel to the average flow direction .

Gravitational effects can easily be accounted for in the invasion percolation model by mapping the system onto a problem where the capillary threshold values are modified linearly by the hydrostatic pressure difference between the two fluids (Wilkinson, 1984; Birovljev et al., 1991; Auradou et al., 1999). By using this theory of percolation in a gradient, Wilkinson (1984) predicted theoretically the scaling of the front width $\xi$ observed in a gravitational field as

$$\xi/a \propto Bo^{\frac{-\nu}{\nu+1}}, \tag{1}$$

where the Bond number $Bo = \Delta\rho g a^2/\gamma$ is the ratio of the difference of the hydrostatic pressure drops in the two fluids on the length scale of a single pore to the capillary pressure drop, and $\nu$ is the critical exponent associated to correlation length divergence in percolation theory ($\nu = 4/3$ in 2D, Wilkinson (1984) ). Here, $\Delta\rho$ is the density difference between the fluids, $g$ the gravitational acceleration, $a$ the characteristic pore size and $\gamma$ the surface tension between the two fluids. Gravitationally-stabilized fluid fronts occurring during very slow two-dimensional drainage have been studied both experimentally and by computer simulations (Birovljev et al., 1991). The results were found consistent with the theoretical prediction of Wilkinson (1984). When a lighter fluid is injected into a heavier fluid from below (Frette et al., 1992; Birovljev et al., 1995; Wagner et al., 1997), gravitational fingering of the displacing fluid through the displaced fluid occurs; a scaling behavior consistent with Eq. (1) has also been found in this unstable case: the characteristic length scale $\xi$ then corresponds to the width of the unstable gravitational fingers, perpendicularly to the average flow direction. The same simple mapping to invasion percolation as described above can also be performed in the case of slow displacement in a rough fracture joint filled with particles (Auradou et al., 1999).

When comparing systems with different capillary pressure threshold distributions, Eq.(1) needs to be modified. From the phenomenology of the invasion process that we have explained above, it is quite intuitive that a gravity - stabilized front in a porous medium presenting a narrow capillary noise (i.e. a narrow distribution of the capillary thresholds) will give a flatter front than a porous medium with a wide capillary noise. Instead of equation Eq. (1), it has therefore been suggested to take into account a dimensionless fluctuation number $F = \Delta\rho g a/W_t$ , in which capillary fluctuations are accounted for in terms of the width $W_t$ of the capillary pressure distribution (Auradou et al., 1999; Méheust et al., 2002). Experiments to check the dependence

192 of the displacement process on the capillary noise ($W_t$) are difficult, because controlling the
193 distribution of threshold capillary pressures in the medium in a systematic way is not
194 straightforward . These experiments therefore remain to be done.

195

## 3 Capillary and viscous effects

197

198     The crossover between capillary fingering and regimes for which viscous effects are
199 dominant was first studied in the pioneering work of Lenormand (1988). He classified the
200 different flow structures in a phase diagram depending on the viscosity contrast $M = \mu_i/\mu_d$
201 between the fluids and the capillary number $Ca = \Delta P_{\text{visc}}/\Delta P_{\text{cap}}$, which is the ratio of the
202 characteristic viscous pressure drop at the pore scale to the capillary pressure drop. Here $\mu_i$ and
203 $\mu_d$ are the viscosities of the injected and displaced fluid, respectively. From the Darcy equation,
204 the capillary number can be evaluated as

$$Ca = \frac{\Delta P_{\text{visc}}}{\Delta P_{\text{cap}}} = \frac{a \nabla P_{\text{visc}}}{\gamma/a} = \frac{\mu a^2 v}{\kappa \gamma}. \tag{2}$$

206 where $\gamma$ is the surface tension, $a$ the characteristic pore size, $\Delta P_{\text{visc}}$ is the viscous pressure drop
207 at pore scale $a$, evaluated from the viscous pressure gradient $\nabla P_{\text{visc}}$, $\kappa$ the intrinsic permeability
208 of the medium, $v$ the seepage velocity associated to the imposed flow rate of the displaced fluid,
209 $\mu$ the viscosity of the most viscous fluid, and $\gamma/a$ the typical capillary pressure drop accross the
210 interface. Lenormand identified three flow regimes: (i) stable displacement, for which the
211 interface roughness is not larger than one linear pore size, (ii) capillary fingering, which we have
212 discussed in section 2, and (iii) viscous fingering, which occurs when large scale fingers of the displacing fluid develop inside a more viscous defending fluid, resulting in a much faster breakthrough of the displacing fluid. It is important to keep in mind that the observed structures will depend on the length scale considered. For large systems it is therefore not meaningful to talk about a sharp transition in a phase diagram between capillary and viscous fingering, because one will always have both structures present, i.e. capillary fingering on small length scales, and either viscous fingering or stable displacement on large length scales. When the two fluids involved have different viscosities, the viscous pressure drop between two points along the fluid interface will typically be different in the two fluids. This viscosity contrast will produce a change in the capillary pressure along the fluid interface, therefore playing a role similar to that of density contrasts in the presence of a gravitational field (see section 2). At sufficiently large length scales, the difference in viscous pressure drop between the two sides of the interface will become larger than the typical fluctuations in capillary pressure threshold. This means that at sufficiently large length scales, and thus for a sufficient large system, viscous pressure drops, rather than capillary forces associated to random capillary thresholds, determine the most likely invaded pores; consequently, viscous fingering will always dominate at sufficiently large scales when a viscous fluid is injected into another more viscous fluid. At these large scales, and in the absence of a stabilizing gravitational effects, two-dimensional flows exhibit tree-like branched displacement structures with a mass fractal dimension $D_v = 1.6$ (Måløy et al., 1985). The fractal dimension of the front, or growing hull, was found experimentally to be around 1, close to the growing interface dimension in DLA models (Feder; 1988).

Méheust et al. (2002) have introduced a generalized fluctuation number

$$F = \frac{\Delta \rho g a - \dfrac{a \mu v}{\kappa}}{W_t} \qquad (3)$$

which is the ratio of the typical total pressure drop in the fluids over one pore, including both viscous and gravitational pressure drops, to the width of the capillary pressure threshold distribution $W_t$. The experiments by Méheust et al. (2002), identical to those by Birovljev et al. (1991) but performed at larger flow rates and therefore under significant viscous effects, showed that the characteristic width of the rough interface parallel to the macroscopic flow could be characterized with the fluctuation number according to an equation analog to Eq.(1):

$$\xi/a \propto F^{\frac{-\nu}{\nu+1}} \qquad (4)$$

where the exponent $\nu/(1+\nu) = 4/7$ is consistent with percolation theory ($\nu = 4/3$) in 2D. Since the viscous pressure field is not homogeneous like the gravitational field, this result is not obvious. Note that in this case $\xi$ can be interpreted as the length scale at which the sum of the viscous and gravitational pressure drop becomes of the same order of magnitude as the spatial fluctuations of the capillary pressure threshold. In terms of fluid-fluid interface, $\xi$ corresponds to the crossover scale between capillary fingering structures at small scale and the stabilized structure, which is linear (dimension 1) at large scales. When the displacement is large enough for viscous forces to play a role, the fractal dimension typical of viscous fingering structures is also seen at intermediate scales (Méheust et al., 2002). Even in the case where the two fluids involved have the same viscosity, the width of the front was found to be consistent with Eq.(4) (Frette et al.,1997). As observed in the experiments of Frette et al. (1997), the effect of trapping turns out, at least in two dimensions, to be of central importance. The trapped islands result in a decrease in the relative permeability of the invaded fluid, which is equivalent to having a fluid with a higher viscosity (as was shown by Frette et al. (1997) by the comparison to simulations allowing trapping or not, with growth along the whole external perimeter of the invader or restricted to the hull). This is the effect responsible for the well-known decreasing dependence of a soil's matric potential on its water content. This result is consistent with the scaling relation Eq.(4), which is expected from theoretical arguments for percolation in a stabilizing gradient (Xu et al., 1998; Wilkinson, 1984; Lenormand, 1989). Note that other scaling relations have been derived theoretically and observed experimentally by other authors, as reported by Wu et al. (1998). As mentioned previously, when a viscous fluid is injected into a 2D medium filled with a more viscous fluid, viscous fingers occur. The scaling of the finger width was studied experimentally by Løvoll et al. (2004) and Toussaint et al. (2005). The measurements were found to be consistent with a scaling law in the form

$$\xi/a \propto \text{Ca}^{-1} \tag{5}$$

This result is different from the scaling laws that can be explained from the theory of percolation in a gradient, as observed from stabilizing viscous or gravitational forces. In the experiments of Løvoll et al. (2004) and Toussaint et al. (2005), contrarily to what happens in a pressure field arising from gravitational effects (as in Méheust et al. (2002)), the viscous pressure field is highly inhomogeneous, constant in front of the fingers, and screened by the fingers behind the invasion front (i.e. the pressure gradient concentrates around the finger tips and decays behind the finger tips, in the stagnant zones). This may explain why the behaviour expected from percolation in a gradient is not observed, but a rather simpler one instead. Note however that scaling laws based on the theory of percolation in a gradient are still expected for some types of unstable flows (Xu et al. 1998).

The scaling law, Eq. (5), can be explained from a simple mean field argument: consider an approximation for the pressure field for which the pressure gradient $\nabla P$ is homogeneous around the mobile sites at the boundary between the two fluids. Let us consider two of these sites, separated by a distance $l$ along the direction of $\nabla P$; the difference between the drops in 281 viscous pressure across the interface at the two sites is $l\nabla P \approx l\text{Ca}\gamma/a^2$. This relation holds not
282 only when the viscosity of one fluid can be neglected with respect to that of the other one (for
283 example, for air and water), in which case a non-zero viscous pressure difference between the
284 two sites occurs in the defending liquid only, and the definition of the capillary number is given
285 by Eq. (2). It also holds in the general case of two viscous fluids, in which case the capillary
286 number can be defined from Eq. (2) by replacing the viscous pressure difference by the
287 difference in the viscous pressure drop across the interface at the two sites, or equivalently by the
288 differential viscous pressure drop $|\Delta P_v^{(i)} - \Delta P_v^{(d)}|$, where the $|\Delta P_v|$ are viscous pressures difference
289 between the two sites in each of the phases, the superscript $^{(i)}$ and $^{(d)}$ denoting respectively the
290 invading- and defending- phase. If the differential viscous pressure difference between these two
291 sites exceeds the characteristic random fluctuations of the capillary pressure threshold from one
292 pore throat to another along the interface, then the viscous pressure field is dominant in
293 determining at which of the two points considered new pores are going to be invaded. On the
294 contrary, if the random fluctuations of capillary threshold exceed the differential viscous
295 pressure drop between the two points, then this random pressure difference component
296 dominates. Assuming that its magnitude $W_t$ is of the same order as the average capillary pressure
297 value, $\gamma/a$ we conclude that capillary effects are expected to dominate for scales $l$ such that
298 $l\text{Ca}\gamma/a^2 < \gamma/a$, whereas viscous effects will dominate for larger scales, such that $l\text{Ca}\gamma/a^2 > \gamma/a$:
299 this explains the observed cross over scale $\xi = a/\text{Ca}$ between the structures characteristic of
300 capillary fingering and those characteristic of viscous fingering. Using a pore-scale simulation
301 where they can tune capillary noise, Holtzman and Juanes (2010) determined a phase diagram of
302 the displacement regime (capillary or viscous fingering) as a function of capillary noise and
303 capillary number, and observed the same crossover, which they explained in a similar manner,

only expressing the prefactor for the scaling law of the crossover scale in terms of typical capillary pressure threshold fluctuations rather than in terms of the mean capillary pressure threshold (that is, they did not assume that $W_t \approx \gamma/a$).

## 4 An example of upscaling from capillary to viscous fingering

In many situations of non-miscible biphasic flow in porous media, the flow gets organized in fingering structures (preferential paths); in these unstable configurations, the fluids arrange in fractal geometries with nontrivial fractal dimensions depending on the observation scale, and the scale range over which each dimension is observed depends on the imposed boundary conditions (such as the macroscopic fluxes). This microscopic structuration has far reaching consequences for the upscaled relationships between, for example, saturation and the pressure difference between the two phases.

An example of such a situation has been mentioned briefly earlier in this review: if gravity is negligible, only capillarity and viscosity play a role on the flow; when a fluid of lower viscosity displaces a more viscous one, the fluid-fluid interface is unstable due to viscous pressure gradients increasing at the most advanced parts of the invader, so that fingers naturally arise. The development of such an interface instability occuring during drainage was studied optically in Hele Shaw cells, at several controlled injection rates (Løvoll et al.2004; Toussaint et al.2005). Air was injected into an artifical porous medium composed of a monolayer of immobile glass beads sandwiched between two glass plates, and initially filled with a wetting dyed glycerol-water solution. The Hele-Shaw cell dimensions were denoted $L \times W \times H$, $H = 1$ mm being the cell thickness as well as the typical glass bead diameter; flow was imposed along the length $L$, with impermeable lateral boundaries defining a channel of width $W$ (see Fig. 1). An occupancy (also termed occupation probability) was defined in the reference frame moving at the average finger speed between the system boundary; in that referential, the invasion structure is seen as a finger fluctuating during the experiment behind its stationary tip: for each point in this reference frame, the proportion of the time where this point is occupied by the invading fluid is a measure of the occupancy. It was shown in these experiments that the pathway of the air, defined as the locations where the occupancy probability exceeds half its maximum value, was a finger of width $\lambda W$ positioned in the centre of the channel, with $\lambda = 0.4$. This was attributed to a similarity between the process of selection of the pore throats to be invaded and a Dielectric Breakdown Model of exponent 2 (Niemeyer et al., 1984), that is, a growth process in which the growth velocity is proportional to the gradient of the driving effect to the power of 2. The presence of a disorder in capillary threshold turns out to be important to enforce a boundary condition analogous to a growth probability along the invader proportional to $[(P-P_{air})/a]^2$, with a field P satisfying Laplace equation due to mass conservation: This result was justified theoretically by computing the average invasion speed taking into account the width of the capillary threshold distribution (Toussaint et al.,2005). This type of invasion structure is illustrated in Fig. 1.

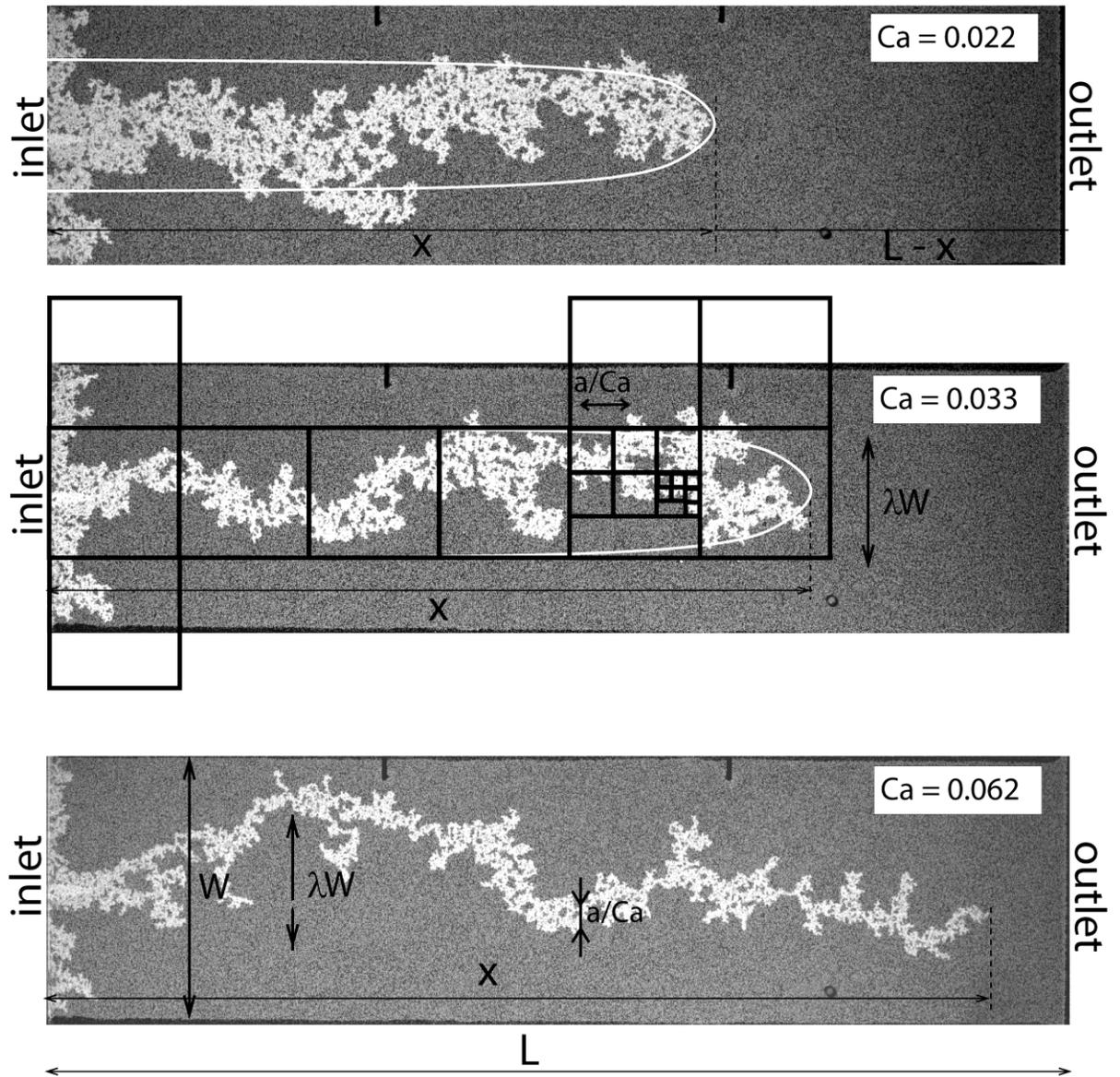

**Figure 1:** Invasion structure of a fluid with a low viscosity (white) into a much more viscous one (dark grey) during drainage in an artificial 2D porous medium of width $W$ and extent $L$, at three different extraction speeds. The position of the invasion tip is denoted $x$. Characteristic crossover scales between fractal regimes, $\lambda W$ and $a/Ca$, separate a straight finger structure, a viscous fingering geometry, and a capillary fingering geometry, down to the pore scale $a$. The black square of various dimensions in the central figure illustrate the types of boxes used in the box counting measure of the fractal dimension: for boxes of a certain side size $l$, one counts the number of boxes $N(l)$ needed to cover the structure. This is done for various sizes, from system size down to pixel size. The scaling of this number as function of the size, $N(l) \sim l^D$ defines the mass fractal dimension $D$. The sizes $W$ and $a/Ca$ turn out to be the limits of scale-ranges with well defined fractal dimensions: $D = 1.00$ above W, $D_v = 1.60$ between W and a/Ca, and $D_c = 1.83$ below. Modified from Løvoll et al. (2011).

358      From approximations on the shape of the pressure around this finger, mostly controlled
359 by the viscous pressure drop, one can derive an upscaled pressure-saturation relation (Løvoll et
360 al.2010).
361      Indeed, the pressure presents to first order a linear viscous pressure drop from the tip of
362 the invasion cluster, at position $x$, to the outlet of the system, at position $L$. Over the rest of the
363 system, the pressure gradient is screened by the finger, rendering the pressure in the wetting fluid
364 essentially constant at a value close to the sum of the air pressure and the entrance pressure $\gamma/a$.
365 Hence, the pressure difference between the two phases, with a pressure in the wetting fluid $P_w$
366 measured at the outlet, the one in the non wetting phase equal to the atmospheric pressure $P_{n.w.}$,
367 and a correction to this entrance pressure, writes as

368 $$\Delta P^* = P_{n.w.} - P_{w.} - \gamma/a = (L-x)\nabla P \quad (6)$$

369 $$= (L-x)\Delta P_{visc}/a \quad (7)$$

370 $$= (L-x)\Delta P_{cap}Ca/a \quad (8)$$

371 $$= (L-x)\frac{\gamma Ca}{a^2} \quad (9)$$

372 where $\Delta P_{visc}$ and $\Delta P_{cap}$ are considered at the pore scale. Thus, there is a linear relationship
373 between the viscous pressure drop accross the cell and the distance between the finger tip and the
374 outlet.
375      In addition, the relationship between saturation and capillary number can be inferred
376 from the fractal structure of the non-wetting invading fluid. At scales above the width $\lambda W$, the
377 finger is a linear structure of dimension 1. Between the scale $\lambda W$ and $a/Ca$, the structure has a
378 viscous fingering geometry of fractal dimension $D_v = 1.60$ (Måløy et al.,1985; Løvoll et
379 al.,2004). Between the crossover scale $a/Ca$ and the pore scale $a$, the structure has a capillary

380  fingering geometry of fractal dimension $D_c = 1.83$. Hence, the total number of pores invaded by
381  the non viscous fluid (n.v.f.) can be evaluated as a function of these fractal dimensions, the ratio
382  of the finger length to its width, $x/(\lambda W)$, and the ratios of the latter length to the two others
383  lengths, the crossover length and the pore size. This leads to:

$$N_{n.w.} = \frac{x}{\lambda W}\left(\frac{\lambda W}{a\text{Ca}^{-1}}\right)^{D_v}\left(\frac{a\text{Ca}^{-1}}{a}\right)^{D_c} \qquad (10)$$

386  Together with the relationship between the total number of pores and the characteristic
387  model dimensions,

$$N_{tot} = \frac{LW}{a^2} \qquad (11)$$

389  and the relationship between the wetting phase and non-wetting phase saturation $S_{n.w.}$,

$$S_{n.w.} = 1 - S_{w.} = \frac{N_{n.w.}}{N_{tot}}, \qquad (12)$$

391  Eq. (10) leads to

$$1 - S_{w.} = S_{n.w.} = \lambda^{D_v - 1}\left(\frac{a}{W}\right)^{2-D_v}\text{Ca}^{D_v - D_c}\left(1 - \frac{a^2 \Delta P^*}{\gamma L \text{Ca}}\right) \qquad (13)$$

395  This relationship allows to collapse all the pressure difference curves measured as a
396  function of saturation in the set of experiments performed by Løvoll et al. (2010) onto a unique
397  master curve for capillary numbers ranging from around 0.008 to 0.12. Note that the rescaled
398  pressure $P^* = a^2 \Delta P^*/(\gamma L \text{Ca}) = (\Delta P^*/M)/(\Delta P_{visc}/a)$ is simply the ratio of the gradient in viscous
399  pressure defined at the scale of the model to the gradient in viscous pressure defined at the pore scale. Apart from the normalization by $\Delta P_{visc}/a$, it is nothing else than what is usually defined at the scale $L$ of the experimental model as the capillary pressure. In other words, Eq. (13) defines the dependence on Darcy/seepage velocity of what is commonly denoted as dynamic capillary pressure, measured at scale $L$. This example shows how both viscous and capillary effects play a role in constraining the geometry of the invasion structures, resulting in a dynamic capillary pressure, as it is traditionally called (Hassanizadeh et al., 2002), that is simply due to the upscaling of the invasion structure, with only capillary and viscous effects seen at the REV scale, and without any dynamic capillary/wetting effects occuring at the pore scale.

Figures 2 and 3 illustrate respectively the raw measurements at several injection speeds and how Eq. (13) allows to collapse these curves of saturation versus pressure: with a reduced saturation $S^* = \lambda^{1-D_v}(a/W)^{-2+D_v} \text{Ca}^{D_c-D_v} S_{n.w.}$ and the rescaled pressure $P' = a^2 \Delta P^*/(\gamma L \text{Ca}) = (\Delta P^*/M)/(\Delta P_{visc}/a)$, Eq. (13) predicts that $P' = 1 - S^*$, which is the theoretical straight line in Fig. 3. This is well followed by the experimental data collapse.

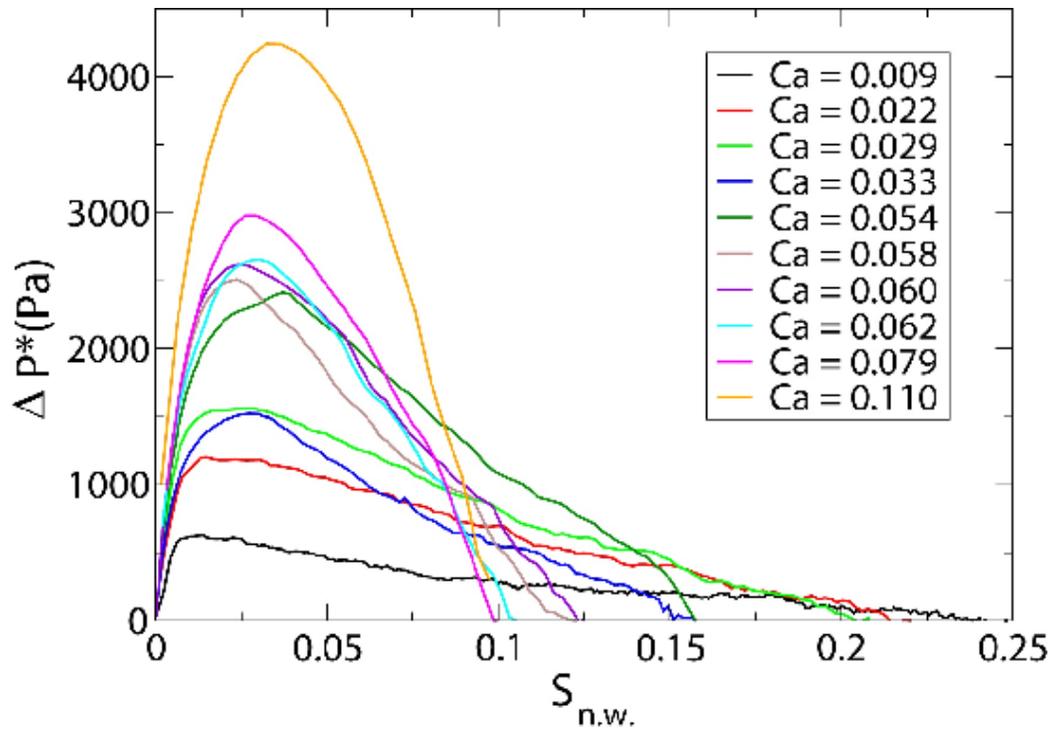

Figure 2: Dependence of the pressure difference between the two phases and the saturation of the invading fluid after removal of the average capillary pressure drop, $\Delta P^* = P_{n.w.} - P_{w.} - \gamma/a$, at different injection speeds. Adapted from Løvoll et al. (2011).

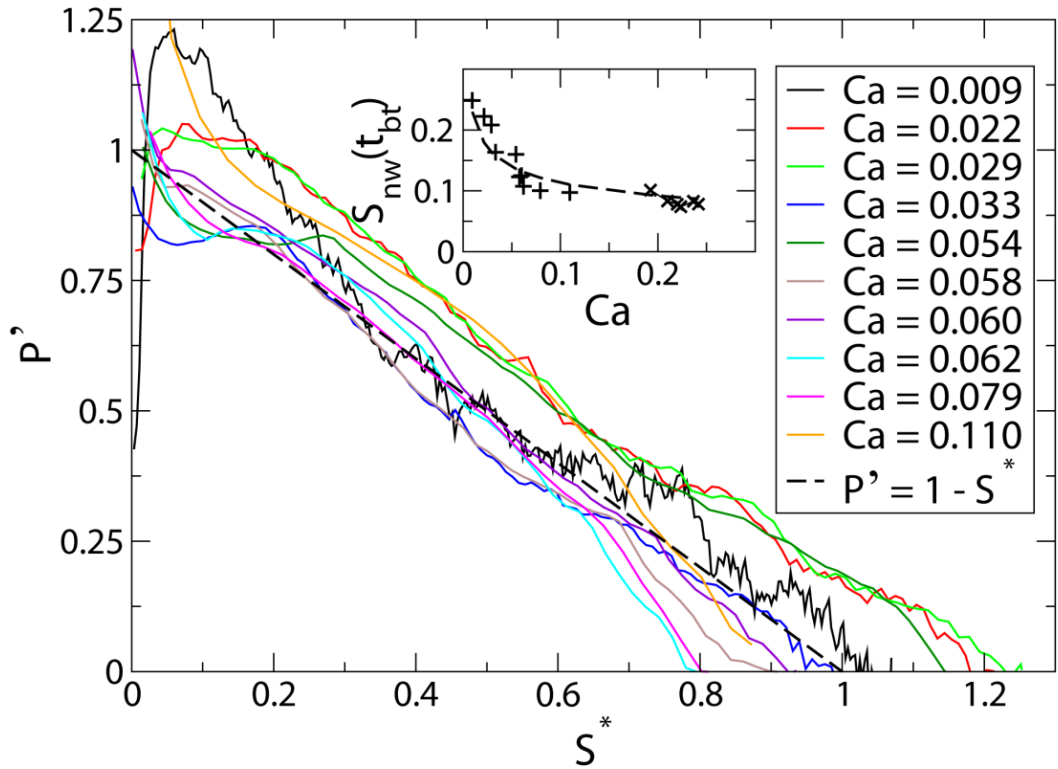

**Figure 3:** The collapse of the relationship between the reduced pressure difference (between the two phases), $P' = a^2 \Delta P^*/(\gamma L \text{Ca})$, and the reduced saturation of the invading fluid, $S^* = \lambda^{1-D_v}(a/W)^{-2+D_v}\text{Ca}^{D_c-D_v}S_{n.w.}$, at different injection speeds, shows the influence of the structure on the upscaling. Note that plots corresponding to lower Ca values have been rescaled more and therefore appear more noisy. Dashed curve: prediction. Inset: Residual saturation at breakthrough. Adapted from Løvoll et al. (2011).

The viscous pressure drop accross the cell drops linearly as the finger progresses into the cell, from a maximum value at the beginning of the invasion of $\gamma L\text{Ca}/a^2$, down to 0 at breakthrough of the invasion finger. In the previous equation, the wetting saturation is indeed initially 1 as it should be at initial total saturation, but we also obtain the final and maximum

430    value of the residual wetting saturation as

431    $$1 - S_{w.r.} = S_{n.w.r.} = \lambda^{D_v - 1} \left(\frac{a}{W}\right)^{2 - D_v} \text{Ca}^{D_v - D_c}. \tag{14}$$

432    This relation between the residual saturation and the capillary number is indeed consistent with

433    the observed residual saturations, as shown in the inset of Fig.3.

434    Besides the relation between saturation and the macroscopic pressure difference between

435    the phases, other macroscopic relations can be obtained via upscaling, as e.g., in some situations,

436    the relative permeability. For example, in other experiments where both fluids were injected at

437    the same time, with both drainage and imbibition happening in the flow simultaneously, the

438    trapped structures of wetting fluids were observed to be fractal up to a certain cutoff depending

439    on the imposed flux (Tallakstad et al., 2009a; 2009b). A scaling law was observed for the

440    relative permeability of wetting viscous fluid in these experiments, with an observed dependence

441    on the imposed flux that can be expressed as $\kappa_{\text{rel}} \propto \text{Ca}^{-1/2}$. The upscaling explaining the cutoff

442    and the structures allowed to explain this measured scaling law.

443    For the pure imbibition case, the macroscopic capillary pressure also presents a dynamic

444    dependence: it was found by Stokes et al. (1986) and Weitz et al. (1987) that the fingering also

445    occurs when the lower viscous fluid displaces the more viscous one, and that the finger width

446    scales as $\text{Ca}^{-1/2}$, a result that is still largely unexplained.

447

448    **5 Conclusion**

449

450    We have discussed the local flow structures that are observed experimentally during

451    drainage in a disordered porous medium. When the viscosity constrast between the two fluids is

452  high (as for air displacing water), the flow structures are fractal, with a fractal dimension that
453  depends on the observation scale. At small scales, capillary fingering exhibits a fractal dimension
454  of 1.8 for two-dimensional media, and between 2 and 2.6 for three-dimensional media. At larger
455  scales a branched structure characteristic of viscous fingering is seen, with a fractal dimension
456  1.6 for two-dimensional systems. The crossover between the two behaviors occurs at a length
457  scale for which the differential viscous pressure drop equals the typical capillary pressure
458  threshold in the medium. This means that for horizontal flow, unstable viscous fingering is
459  always seen at large enough scales, even if the medium exhibits no permeability heterogeneities
460  at the Darcy scale. From the definition of the crossover length, it follows that it scales as the
461  inverse of the capillary number, which explains why experiments performed at a given
462  experimental scale and at very slow flow rates have evidenced capillary fingering, while those
463  performed at very large flow rates have evidenced viscous fingering. As for the effect of gravity,
464  it can be to either destabilize or stabilize the interface, depending on which fluid is the densest.
465  In the latter case, it acts against capillary effects and, when the displacing fluid is the most
466  viscous, against the destabilizing viscous forces, resulting in an amplitude of the interface
467  roughness that scales as a power law of the generalized fluctuation number (or generalized Bond
468  number $Bo - Ca$, as introduced by Méheust et al. (2002)). In horizontal two-dimensional flows,
469  viscous fingering is observed to occur up to another characteristic length that is a fixed fraction
470  of the width of the medium. Upscaling of the local flow structures is possible once one knows
471  the fractal dimensions typical of the flow regimes, and the relevant length scale range for each of
472  them. We have given an example of how the measured capillary pressure can be related
473  theoretically to water saturation, a relation that is confirmed by measurements. In that example,
474  the capillary pressure measured at the scale of the experimental setup exhibits dynamic features,

i. e., a dependence on the flow rate, that is fully explained by the geometry of the upscaling, without any dynamic effects in the physical capillary pressure as defined at the pore/interface scale.

## 6 Acknowledgements

This work was supported by the CNRS through a french-norwegian PICS grant, the Alsace region through the REALISE program, and the Norwegian NFR.

670

671      Figure Captions:

672      **Figure 1:** Invasion structure of a fluid with a low viscosity (white) into a much more
673 viscous one (dark grey) during drainage in an artificial 2D porous medium of width $W$ and
674 extent $L$, at three different extraction speeds. The position of the invasion tip is denoted $x$.
675 Characteristic crossover scales between fractal regimes, $\lambda W$ and $a/Ca$, separate a straight
676 finger structure, a viscous fingering geometry, and a capillary fingering geometry, down to the
677 pore scale $a$. The black square of various dimensions in the central figure illustrate the types of
678 boxes used in the box counting measure of the fractal dimension: for boxes of a certain side size
679 $l$, one counts the number of boxes $N(l)$ needed to cover the structure. This is done for various
680 sizes, from system size down to pixel size. The scaling of this number as function of the size,
681 $N(l) \sim l^D$ defines the mass fractal dimension $D$. The sizes $W$ and $a/Ca$ turn out to be the limits of
682 scale-ranges with well defined fractal dimensions: $D = 1.00$ above W, $D_v = 1.60$ between W
683 and a/Ca, and $D_c = 1.83$ below. Modified from Løvoll et al. (2011).
684

685      **Figure 2:** Dependence of the pressure difference between the two phases and the
686 saturation of the invading fluid after removal of the average capillary pressure drop,
687 $\Delta P^* = P_{n.w.} - P_{w.} - \gamma/a$, at different injection speeds. Adapted from Løvoll et al. (2011).
688

689      **Figure 3:** The collapse of the relationship between the reduced pressure difference

690   (between the two phases), $P' = a^2 \Delta P^* / (\gamma L \text{Ca})$, and the reduced saturation of the invading fluid,
691   $S^* = \lambda^{1-D_v} (a/W)^{-2+D_v} \text{Ca}^{D_c - D_v} S_{n.w.}$, at different injection speeds, shows the influence of the
692   structure on the upscaling. Note that plots corresponding to lower Ca values have been rescaled
693   more and therefore appear more noisy. Dashed curve: prediction. Inset: Residual saturation at
694   breakthrough. Adapted from Løvoll et al. (2011).
695